# How to democratize Internet of Things devices.
# A participatory design research

Matteo Zallio[1,*], John McGrory[2], and Damon Berry[2]

[1] Stanford University, Autonomous Systems Laboratory, Lomita Mall, Stanford, CA, USA
{matteo.zallio}@Stanford.edu
[2] Technological University Dublin, School of Electrical & Electronic Engineering, Dublin, Ireland
{john.mcgrory, damon.berry}@TuDublin.ie

**Abstract.** The global introduction of affordable Internet of Things (IoT) devices offers an opportunity to empower a variety of users with different needs.
However, many off-the-shelf digital products are still not widely adopted by people who are hesitant technology users or by older adults, notwithstanding that the design and user-interaction of these devices is recognized to be user-friendly. Considering the habits and preferences of those who have used mainly analog devices for most of their lives, such users may encounter challenges like digital illiteracy and technology anxiety when faced with newly-released digital IoT-based products.
In view of the potential of IoT devices, how can we reduce obstacles of a cohort with low digital literacy and enable them to be equal participants in the digitalized world?
This article shows the method and results achieved in a community-stakeholder workshop, developed through the participatory design research methodology, aiming at brainstorming problems and scenarios through a focus group and a structured survey. The research activity focused on (1) understanding pain points in the human-machine interaction process, (2) brainstorming scenarios to increase the usability of off-the-shelf IoT devices for hesitant users and (3) investigating strategies to improve digital literacy and reduce technology anxiety.
A notable result was a series of feedback items pointing to the importance of facilitating educational experiences through learning resources in order to support individuals with different abilities, age, gender expression, to better adopt off-the-shelf IoT solutions.
This first User Experience research generated considerations for designing open-source learning tools and generating interaction proposals to foster inclusive and more accessible use of IoT technologies for improving older people's daily activities.

**Keywords:** Participatory Design · User Experience Research · Usability · Internet of Things · Smart Ageing Friendly Environments · Inclusive Design · Learning Tools

## 1. Introduction

The ageing profiles of populations across the globe, from Europe to USA, to Asia, have consequences for different aspects of daily life [1].
The United Nations, with the Sustainable Development Goals (UN-SDG) [2] describe how health & well-being, sustainable cities & communities, innovation & infrastructure, represent a baseline for guaranteeing a better future for these populations.
These goals affect the use of different devices, services and solutions as well as the settings where humans spend most of their life: in the community, at work, and at home.

According to the World Health Organization [3], sustainable environments provide effective support for well-being & socialization. However, every user has a distinct sense of place [4] as well as different ways of interacting with the surrounding environments, objects, technology and to socialize and connect with others. Technologies, in particular Internet of Things (IoT) devices, collaborative robotic applications and cloud-based systems have been released into the consumer market over the past decade. Such devices have a potential to foster independent living, to increase psychophysical health and advance older adults' physical and social activities [5, 6].

Inexperienced or tentative users may encounter challenges such as digital illiteracy and technology anxiety with newly released digital IoT products notwithstanding that the configuration (e.g. Plug & Play), design and user-interaction of these devices is considered to be user-friendly [7, 8, 9].

These challenges can be related to concerns about tolerance of error, system reliability, social exclusion burden, affordability, security and privacy of data generated about the users and their dwelling, and higher energy consumption [10, 11, 12].

The question arises: given the potential of IoT devices, how can we reduce obstacles of a cohort with low digital literacy and enable marginalized users to fully join the digitalized world?

The proposed research aims to investigate the factors that could lead to improvements in the use of off-the-shelf IoT devices for certain individuals through the participatory design research methodology. Initial discussions with international experts in disciplines such as product inclusion, technology development and caregiving services were activated as a starting point for understanding the impact of learning resources and discuss them in a community-stakeholder workshop.

The main goals of this study were to (1) understand pain points in the human-machine interaction process by involving experts working with older individuals in conversations and ethnographic research, (2) stimulate brainstorming of scenarios to increase the usability of off-the-shelf IoT devices for hesitant users, and (3) identify learning strategies that could lead to reduction of digital illiteracy barriers and improve confidence in using IoT-based technologies.

## 2. Materials and Methodology

The impetus for investigating digital literacy and awareness of older adults in the use of IoT-based technologies arose from several years of research with stakeholders, as well as from published academic reports [13, 14].

The methodology chosen to perform the initial exploratory research focused on the participatory design research methodology. Participatory design is a democratic process for understanding and designing systems involving human work, based on the argument that people should be involved in the designs process, and that all stakeholders have equal opportunities to provide inputs into this process [15, 16].

According to Spinuzzi, participatory design is research [17]. Participatory research is characterized by defining problems, ideas addressing problems and solutions in cooperation with and for people. Designers and researchers see themselves as facilitators who attempt to empower users in understanding their problems and framing ideas to solve needs [18].

The main activity of this participatory research was framed with a workshop, composed by a focus group and a structured survey, deployed by conducting the participants (with already extensive working experience with senior citizens) in the mindset of older users, replicating some of their obstacles as a low digital literacy cohort.

The focus group procedure, which offers a deep collection and examination of detailed human experiences [19], was undertaken with the goal of analyzing views and perspectives regarding: (a) generalized causes that illuminate problems associated with newly released off-the-shelf IoT devices, (b) understand what specific pain points people may encounter when using a voice assistant or a smart plug in conjunction with certain Activities of Daily Living (ADLs) and Instrumental Activities of Daily Living (IADLs), (c) brainstorm ideas on how those devices might be used to solve different key challenges and which aspects might be difficult to learn by different users.

The structured survey [20] was used to obtain final remarks and feedback on the previously brainstormed ideas and to understand future strategies to develop learning tools for empowering older individuals to benefit from different IoT technologies. Those two complementary tools were jointly selected to enable a deeper understanding of stakeholders' views, that would otherwise be hard to achieve with a single quantitative research method. The intention of this combination was to correlate data from the focus group and survey in order to assure consistency of information and to assure anonymity and confidentiality of participants (with the structured survey) by collecting insights that would be otherwise difficult to collect only with a focus group.

### 2.1 Recruitment of Participants

The community-stakeholder workshop was included in the activities of the International Forum on Active Assisted Living, held in Aarhus, Denmark on September 2019. A total amount of 45 participants, coming from several European countries, with different linguistic background, aged above 25 years old chose to participate. The group included healthcare sector specialists, design and engineering professionals and academic experts, all experienced working closely to older adults, as well as a cohort of senior people. Participants voluntary participated to a two-hour workshop session.

The activity was articulated within a two-layer participatory design research activity: a demonstration session, following a brainstorming of problem causes (focus group), and feedback evaluation (structured survey).

An introductory twenty-minute demonstration session on constraints and difficulties in learning how to use certain IoT devices (in order to enable participants to walk in older people's shoes) was provided by session facilitators, with two examples for voice assistants, smart lighting systems and wearable devices. As participation policy, voluntary subjects didn't receive any compensation for their participation and were able to drop the session at any time.

### 2.2 Data Collection

After the introductory session, participants were divided in to twelve working groups of three to four individuals each and were asked to define causes of problems regarding the use of IoT-based devices. The focus group activity was framed around two selected off-the-shelf IoT technologies: a voice assistant and a smart plug device.

In order to facilitate the participatory activity, each group received an infographic card (as shown in Figure 1) where a series of ADLs and IADLs icons were shown to facilitate problems and use cases brainstorming. Each group was equipped with a whiteboard, post-its and pens to crystalize and note thoughts and statements through text, icons and symbols.

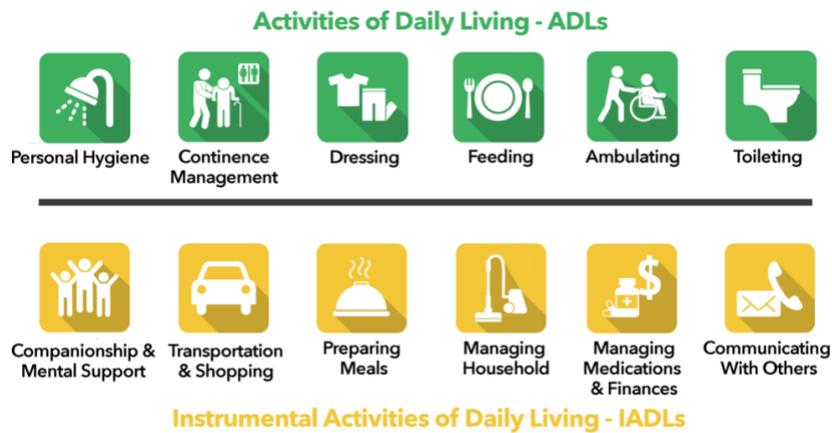

**Figure 1.** Infographic representation of main ADLs and IADLs to facilitate the understanding of usage scenarios of IoT technologies: voice assistant device and a smart plug.

Mentorship from the research team was delivered to facilitate stakeholders in expressing as many feedback as possible and prototyping scenarios of use of the selected devices. Conceptualizing ideas was driven by the paradigm of quantity over quality in order to emphasize the understanding of certain features, use and operation of a voice assistant and a smart plug for people who have limited literacy skills.

In the second stage, participants were asked to provide feedback through a structured survey in order to clarify what type of learning resources would be useful to better interact with IoT devices, what sort of extended audience could benefit from those learning resources and how should those tools be developed and shared across the community.

## 3. Participatory Design Research

Out of twelve groups, respectively six provided feedback regarding the voice assistant device and six regarding the smart plug. The large quantity of information was condensed into two infographics (Figure 2, 3) where ADLs and IADLs were used as support criteria to facilitate the brainstorming activity.

### 3.1 Focus Group Results and Feedback: Voice Assistant

The researchers noticed throughout the session that stakeholders were pretty familiar with voice assistant devices and their knowledge of use and interaction with those IoT technologies was fair. However, the majority of participants expressed concerns regarding ease-of-use for older people. In virtually all six groups, participants created simple use cases for both ADLs and IADLs and experimented with verbal interactions in order to simulate a real-life context.

Some of the most prominent and common feedback for the use of voice-based assistance in relation to ADLs were.
1. "Receive step-by-step instructions and support with morning routine/orientation for people with early stage dementia, by considering a question/answer procedure to be facilitated by the device."
2. "Improve the conversation between end-users and family members to keep an active lifestyle; for example: 'you need to stand up', 'you need to walk', 'it's time for medication' or 'it's time for a tea with friends'"
3. "Serve as a home companion for considering different clothing tips according to weather forecast, trends in the market or seasonal preferred choices of the community."

Interestingly, in the field of IADLs, more curiosity, awareness and creative feedback was distilled from the focus group, enabling different interactions with other devices, family members and caregivers out of the home environment.
1. "Support meal preparation, suggestions for different diets, vegetarian/vegan options and improve creativity in cooking, and as a result satisfaction in meal preparation or ordering from a local shop."
2. "Improve communications of certain topics among peers but also with chatbots that would lower the barrier and stigma of restricted or taboo topics."
3. "Suggest guidance using simple chains of instructions for learning how to improve DIY skills and small repairs at home."
4. "Schedule time for personal activities, housekeeping, and connect other personal devices to previously scheduled activities."

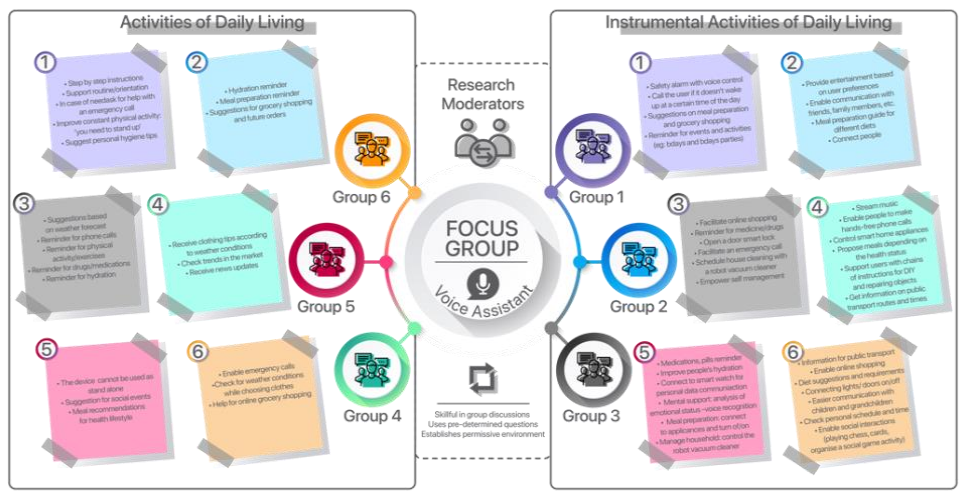

**Figure 2.** Infographic with representation of feedback elicited from the focus group on voice assistant device in the context of use with ADLs (on the left) and IADLs (on the right).

Some of the most prominent and representative quotes collected during this research activity related mainly to three usability aspects: a) device reliability: use of sentences by the user that are not currently understandable by voice assistance devices, due to limitations of the Digital Audio Processing software; b) tolerance of error and easiness of deleting an input that was erroneously initiated; c) lack of trust about the storage, processing and privacy of the audio registrations, (during various activities), in particular personal and intimate conversations in environments such as bathroom and bedroom.

### 3.2 Focus Group Results and Feedback: Smart Plug

The second half of participants focused mainly on understating scenarios associated with use of a smart plug that upgrades traditional home appliances in to connected, re-motely controlled devices.
The majority of participants were not fully familiar with the potential offered by smart plugs, so workshop facilitators had to emphasize several functions during the initial explanatory session. For example, a lamp connected to a smart plug is controllable from smartphones or tablets and the switch on the lamp can be left unused.
As a result, many stakeholders developed several creative scenarios of use regarding ADLs and IADLs and expressed major concerns related to economic affordability, possible higher energy consumption and system reliability. Major feedback regarding the ADLs scenario focused mostly on feeding people and personal hygiene.
1. "Regularly control the use of personal electronic devices in the bathroom to detect usage patterns for personal hygiene."
2. "Detect energy consumption of kitchen appliances (refrigerator, kettle, oven, cooker) for detecting anomalies and misuse."
3. "Check, through remote control of switches, the activities performed by users, and if needed, set up notifications or alarms."
4. "Create ambient to follow daily routine."

Focusing on IADLs activities such as companionship and social support, managing household and communication with others were the most discussed by stakeholders.
1. "Find out easily how to control different devices, the energy consumption and schedule timing and zone activity in a reliable, user-friendly way."
2. "Understand usage patterns of appliances in the living room, kitchen, bedroom, bathroom and set up requests and alerts (via mail, texts, calls) primary to home users, then to family members and caregivers."
3. "Connect smart plugs with web-based services to foster productivity at home by connecting sensors and actuators that control devices automatically (according to use patterns)."
4. "Connect and turn off-on devices (TV, phone, tablet, computer, robot companions, cameras) to enable conversation and communication with others."

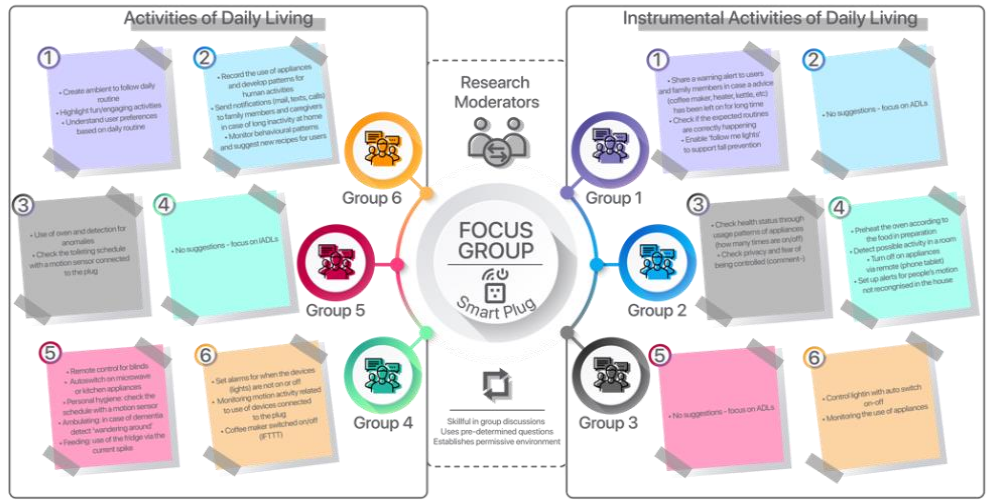

**Figure 3.** Infographic with representation of feedback elicited from the focus group on smart plug device in the context of use with ADLs (on the left) and IADLs (on the right).

The scenario regarding the use of smart plug devices was slightly more complex, mostly because of the inner lack of knowledge of potential functions and capabilities of the selected device.

The cost of the device, compared to a traditional alternative with no Internet capability, was a particular topic of discussion among most of the groups. A further challenge posed by the group regarded the ease-of-use of function integration and the human-computer interaction process.

### 3.3 Structured Survey: Learning tools

During the last stage of participatory design research, a structured on-line survey was submitted to participants. Three questions were submitted to participants with a Likert evaluation scale (1 to 10) to rank which solution should take priority and was most important. Wittink & Bayer [21] noted that a ten-point Likert scale can improve measurement reliability, reduce multicollinearity problems and minimize skewness in the distribution of the data. Clarifications were elicited on awareness-related concerns and around types of learning resource for better understanding the use of IoT-based devices. The concluding survey was answered by a total of 20 of the workshop participants.

The survey was framed across three main questions respectively on: finding modalities to enable users to better understand the use and functionalities of off-the-shelf IoT-based devices, on identifying the major features that learning resources should include and also identifying potential audiences for those learning resources.

Details of the survey questions are listed below:
  Question 1: How do you envision the informative learning tools to be structured?
  Question 2: Who should benefit the learning tool most?
  Question 3: What are the most important features to include in the learning tools?

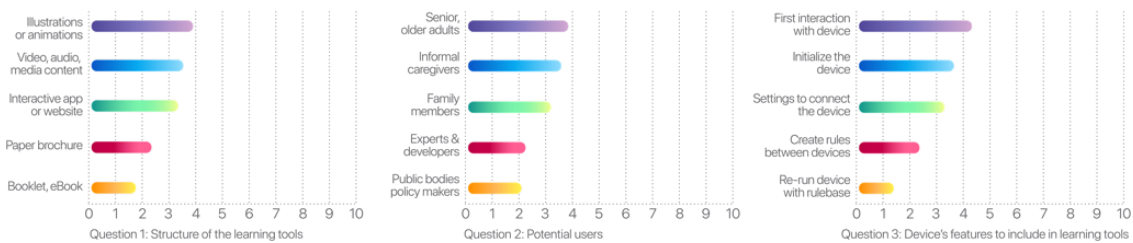

**Figure 4.** Data visualization of the structured survey responses.

Major feedback brought attention to the quality of learning material, delivery of the learning resources and the languages to be used, while a minor relevance was given to the inner topics and the depth of instructions to be included in the learning tools.

## 4. Discussion: Open-Source Learning Tools

The results of the community-stakeholder workshop presented in the previous sections confirmed the relevance of the research question and presented some interesting topics of investigation in relation to digital illiteracy and technology anxiety.

The use of the chosen methodology (participatory design research) allowed the research team to mobilize a significant number of stakeholders with experience in dealing with older adults and elicit meaningful scenarios with constraints and potentialities for the use of certain IoT-based devices.

The focus group supported participants to 'dig deeply' in to different device features, to develop use cases and brainstorm constraints, issues and potentials in relation to daily routine activities.

Discussions allowed participants to frame contributions around system reliability, ability to learn an easy methodology to control the device, familiarization with interface and commands, the justification of higher costs and higher energy consumption compared to a traditional devices and difficulties in understanding privacy and data management of IoT-based technologies.

The participatory activity led stakeholders towards the common need of developing tools for improving digital literacy, such as interactive user manuals, educational resources and informational experiences.

The final results underlined the importance of developing a new form of learning experiences, that would be open-source (supported and created by the community of users), independently developed, with emphasis on different educational levels (from beginners, to experts), and easily accessible by different media (smartphones, tablets, tv, computers, voice assistant appliances, wearable devices).

In order to highlight the major features from the survey, a normalization of ratings by adjusting values measured on different scales to align to a normal distribution was performed.

According to each survey question a detailed visualization of normalized data was created (Figure 5) and major responses underlined the importance to three main aspects:

1. The structure and media to use for developing and communicating the learning resources: graphical illustrations or animations (infographics and icons), video, audio and media content, and interactive mediums (application or website).
2. Enlarge the audience of possible users: from seniors, to informal caregivers, and family members who want to learn efficiently how to interact with a new device or simply receive user-friendly usage tips.
3. The main features that are encountered during a first interaction with the device (turning on-off-standby), initializing and setting up the device (log in-out options, settings, privacy), integration and connection of the device with others (Bluetooth-Wi-Fi pairing, data transferring).

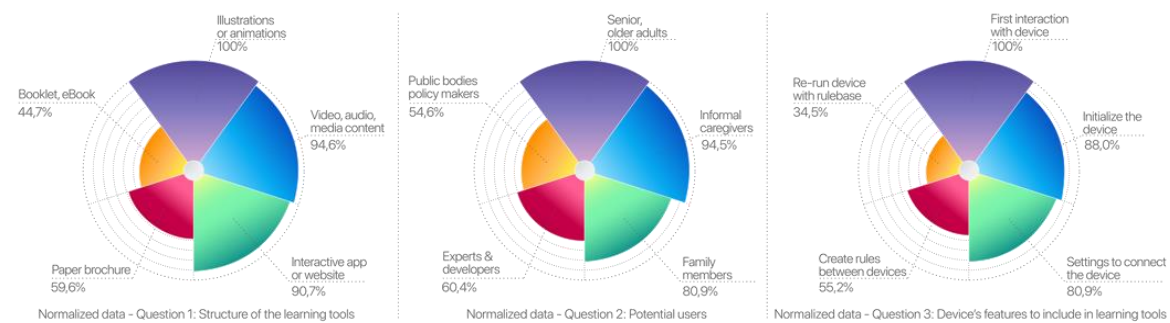

**Figure 5.** Data visualization of the normalized results of the survey responses.

All these criteria are based on the assumption that a large majority of people (90%) aged 55 and more have a Wi-Fi network at home and can easily access to the Internet, computers and smartphones, according to the Linkage Connect Technology Survey [22]. This ability to easily access information allowed workshop participants to be more confident in considering learning tools based on interactive mediums like apps with embedded audio and video recordings and infographic, animated content.

Data interpretation was grounded both in focus group and survey activities and the findings highlighted the relevant need to develop educational tools to foster technology skills and awareness, improve the ability to choose devices according to specific personal challenges, and to increase self-awareness capabilities to interact with a variety of off-the-shelf IoT-based devices.

## 5. Conclusions and future work

This research generated a series of solution-oriented feedback that in the opinion of the authors could be useful to implement the creation of learning experience tools.

The main scope of this work was to develop a user-based research through ethnographic research to offer grounded information for ideating, prototyping and testing learning solutions to ease use of IoT devices across different countries, languages, cultures and individuals. The feedback collected through the participatory design research underlined the importance of developing resources to facilitate educational experiences to support individuals with different abilities, age, gender expression in using new technologies to support daily activities. These educational experiences have the potential to foster the adoption of off-the-shelf IoT devices for improving the performance of daily activities and to develop a better human-machine interaction process.

The next steps of project development will focus on deploying prototypes of training packages to foster learning experiences, based on interactive mediums, with multilingual platforms, through the use of graphical illustrations with a specific goal of enlarging the audience of potential users. The main features that will be included in the learning resources will initially focus, as suggested from consultation, on three steps of human-machine interaction: turning the device on/off, general and privacy settings and technology integration. As outcomes it is foreseen that those resources could lower cultural and language barriers through the inclusive use of off-the-shelf IoT technology, boost social inclusion with the open-source future creation of those learning resources, and in the longer term foster cross-sectoral cooperation among companies, educational facilities and institutions.

**Acknowledgments.** This research, as an impulse into a larger international project, was made possible thanks to the Active Assisted Living Forum, in Aarhus, Denmark on September 23-25, 2019. A team of researchers from Stanford University (USA) and Technological University Dublin (Ireland) developed the participatory design research and successfully run the collaborative workshop supported by the AAL Programme.